\documentstyle[12pt]{article}

\def\beq{\begin{equation}}
\def\eeq{\end{equation}}
\def\eea{\end{eqnarray}}
\def\bq{\begin{quote}}
\def\eq{\end{quote}}

\newcommand{\EQ}{\begin{equation}}
\newcommand{\EN}{\end{equation}}
\newcommand{\bea}{\begin{eqnarray}}
\newcommand{\ena}{\end{eqnarray}}

\renewcommand{\a}{\alpha}
\renewcommand{\b}{\beta}

\renewcommand{\d}{\delta}
\newcommand{\th}{\theta}

\newcommand{\pa}{\partial}

\newcommand{\k}{\kappa}

\renewcommand{\l}{\lambda}
\renewcommand{\L}{\Lambda}
\newcommand{\m}{\mu}

\newcommand{\p}{\pi}

\newcommand{\r}{\rho}

\newcommand{\s}{\sigma}

 \newcommand{ \Xb}{\bar{X}}
 \newcommand{\Db}{\bar{D}}
 
 \newcommand{\Psib}{\bar{\Psi}}

\newcommand{\pp}{\mid \!\!\! =}

\catcode`\@=11
\long\def\@makefntext#1{ 
\protect\noindent \hbox to 3.2pt {\hskip-.9pt
$^{{\ninerm\@thefnmark}}$\hfil}#1\hfill} 

\def\thefootnote{\fnsymbol{footnote}}
 \def\@makefnmark{\hbox to 0pt{$^{\@thefnmark}$\hss}}  

\def\ps@myheadings{\let\@mkboth\@gobbletwo
\def\@oddhead{\hbox{} 
\rightmark\hfil\ninerm\thepage}
\def\@oddfoot{}\def\@evenhead{\ninerm\thepage\hfil 
\leftmark\hbox{}}\def\@evenfoot{}
\def\sectionmark##1{}\def\subsectionmark##1{}}

\begin{document}

\newcommand{\symbolfootnote}{\renewcommand{\thefootnote}
	{\fnsymbol{footnote}}}
\renewcommand{\thefootnote}{\fnsymbol{footnote}}
\newcommand{\alphfootnote}
	{\setcounter{footnote}{0}
	 \renewcommand{\thefootnote}{\sevenrm\alph{footnote}}}

\newcounter{sectionc}\newcounter{subsectionc}\newcounter{subsubsectionc}
\renewcommand{\section}[1] {\vspace{0.6cm}\addtocounter{sectionc}{1}
\setcounter{subsectionc}{0}\setcounter{subsubsectionc}{0}\noindent
	{\bf\thesectionc. #1}\par\vspace{0.4cm}}
\renewcommand{\subsection}[1] {\vspace{0.6cm}\addtocounter{subsectionc}{1}
	\setcounter{subsubsectionc}{0}\noindent
	{\it\thesectionc.\thesubsectionc. #1}\par\vspace{0.4cm}}
\renewcommand{\subsubsection}[1]
{\vspace{0.6cm}\addtocounter{subsubsectionc}{1}
	\noindent {\rm\thesectionc.\thesubsectionc.\thesubsubsectionc.
	#1}\par\vspace{0.4cm}}
\newcommand{\nonumsection}[1] {\vspace{0.6cm}\noindent{\bf #1}
	\par\vspace{0.4cm}}

\newcounter{appendixc}
\newcounter{subappendixc}[appendixc]
\newcounter{subsubappendixc}[subappendixc]
\renewcommand{\thesubappendixc}{\Alph{appendixc}.\arabic{subappendixc}}
\renewcommand{\thesubsubappendixc}
	{\Alph{appendixc}.\arabic{subappendixc}.\arabic{subsubappendixc}}

\renewcommand{\appendix}[1] {\vspace{0.6cm}
        \refstepcounter{appendixc}
        \setcounter{figure}{0}
        \setcounter{table}{0}
        \setcounter{equation}{0}
        \renewcommand{\thefigure}{\Alph{appendixc}.\arabic{figure}}
        \renewcommand{\thetable}{\Alph{appendixc}.\arabic{table}}
        \renewcommand{\theappendixc}{\Alph{appendixc}}
        \renewcommand{\theequation}{\Alph{appendixc}.\arabic{equation}}
        \noindent{\bf Appendix \theappendixc #1}\par\vspace{0.4cm}}
\newcommand{\subappendix}[1] {\vspace{0.6cm}
        \refstepcounter{subappendixc}
        \noindent{\bf Appendix \thesubappendixc. #1}\par\vspace{0.4cm}}
\newcommand{\subsubappendix}[1] {\vspace{0.6cm}
        \refstepcounter{subsubappendixc}
        \noindent{\it Appendix \thesubsubappendixc. #1}
	\par\vspace{0.4cm}}

\def\abstracts#1{{
	\centering{\begin{minipage}{30pc}\tenrm\baselineskip=12pt\noindent
	\centerline{\tenrm ABSTRACT}\vspace{0.3cm}
	\parindent=0pt #1
	\end{minipage} }\par}}

\newcommand{\bibit}{\it}
\newcommand{\bibbf}{\bf}
\renewenvironment{thebibliography}[1]
	{\begin{list}{\arabic{enumi}.}
	{\usecounter{enumi}\setlength{\parsep}{0pt}
\setlength{\leftmargin 1.25cm}{\rightmargin 0pt}
	 \setlength{\itemsep}{0pt} \settowidth
	{\labelwidth}{#1.}\sloppy}}{\end{list}}

\topsep=0in\parsep=0in\itemsep=0in
\parindent=1.5pc

\newcounter{itemlistc}
\newcounter{romanlistc}
\newcounter{alphlistc}
\newcounter{arabiclistc}
\newenvironment{itemlist}
    	{\setcounter{itemlistc}{0}
	 \begin{list}{$\bullet$}
	{\usecounter{itemlistc}
	 \setlength{\parsep}{0pt}
	 \setlength{\itemsep}{0pt}}}{\end{list}}

\newenvironment{romanlist}
	{\setcounter{romanlistc}{0}
	 \begin{list}{$($\roman{romanlistc}$)$}
	{\usecounter{romanlistc}
	 \setlength{\parsep}{0pt}
	 \setlength{\itemsep}{0pt}}}{\end{list}}

\newenvironment{alphlist}
	{\setcounter{alphlistc}{0}
	 \begin{list}{$($\alph{alphlistc}$)$}
	{\usecounter{alphlistc}
	 \setlength{\parsep}{0pt}
	 \setlength{\itemsep}{0pt}}}{\end{list}}

\newenvironment{arabiclist}
	{\setcounter{arabiclistc}{0}
	 \begin{list}{\arabic{arabiclistc}}
	{\usecounter{arabiclistc}
	 \setlength{\parsep}{0pt}
	 \setlength{\itemsep}{0pt}}}{\end{list}}

\newcommand{\fcaption}[1]{
        \refstepcounter{figure}
        \setbox\@tempboxa = \hbox{\tenrm Fig.~\thefigure. #1}
        \ifdim \wd\@tempboxa > 6in
           {\begin{center}
        \parbox{6in}{\tenrm\baselineskip=12pt Fig.~\thefigure. #1 }
            \end{center}}
        \else
             {\begin{center}
             {\tenrm Fig.~\thefigure. #1}
              \end{center}}
        \fi}

\newcommand{\tcaption}[1]{
        \refstepcounter{table}
        \setbox\@tempboxa = \hbox{\tenrm Table~\thetable. #1}
        \ifdim \wd\@tempboxa > 6in
           {\begin{center}
        \parbox{6in}{\tenrm\baselineskip=12pt Table~\thetable. #1 }
            \end{center}}
        \else
             {\begin{center}
             {\tenrm Table~\thetable. #1}
              \end{center}}
        \fi}

\def\@citex[#1]#2{\if@filesw\immediate\write\@auxout
	{\string\citation{#2}}\fi
\def\@citea{}\@cite{\@for\@citeb:=#2\do
	{\@citea\def\@citea{,}\@ifundefined
	{b@\@citeb}{{\bf ?}\@warning
	{Citation `\@citeb' on page \thepage \space undefined}}
	{\csname b@\@citeb\endcsname}}}{#1}}

\newif\if@cghi
\def\cite{\@cghitrue\@ifnextchar [{\@tempswatrue
	\@citex}{\@tempswafalse\@citex[]}}
\def\citelow{\@cghifalse\@ifnextchar [{\@tempswatrue
	\@citex}{\@tempswafalse\@citex[]}}
\def\@cite#1#2{{$\null^{#1}$\if@tempswa\typeout
	{IJCGA warning: optional citation argument
	ignored: `#2'} \fi}}
\newcommand{\citeup}{\cite}

\def\fnm#1{$^{\mbox{\scriptsize #1}}$}
\def\fnt#1#2{\footnotetext{\kern-.3em
	{$^{\mbox{\sevenrm #1}}$}{#2}}}

\font\twelvebf=cmbx10 scaled\magstep 1
\font\twelverm=cmr10 scaled\magstep 1
\font\twelveit=cmti10 scaled\magstep 1
\font\elevenbfit=cmbxti10 scaled\magstephalf
\font\elevenbf=cmbx10 scaled\magstephalf
\font\elevenrm=cmr10 scaled\magstephalf
\font\elevenit=cmti10 scaled\magstephalf
\font\bfit=cmbxti10
\font\tenbf=cmbx10
\font\tenrm=cmr10
\font\tenit=cmti10
\font\ninebf=cmbx9
\font\ninerm=cmr9
\font\nineit=cmti9
\font\eightbf=cmbx8
\font\eightrm=cmr8
\font\eightit=cmti8

\begin{flushright}
IFUM-491-FT\\
BRX-TH-368
\end{flushright}

\centerline{\tenbf  LAGRANGIAN DESCRIPTION OF N=2 MINIMAL MODELS}
\baselineskip=16pt
\centerline{\tenbf  AS CRITICAL POINTS OF LANDAU-GINZBURG THEORIES}
\vspace{0.8cm}
\centerline{\tenrm M.T. GRISARU}
\baselineskip=13pt
\centerline{\tenit  Physics Department, Brandeis University}
\baselineskip=12pt
\centerline{\tenit Waltham, MA 02254, USA}
\vspace{0.3cm}
\centerline{\tenrm and}
\vspace{0.3cm}
\centerline{\tenrm  D. ZANON}
\baselineskip=13pt
\centerline{\tenit  Dipartimento di Fisica dell'Universit$\grave{a}$ di Milano
and INFN,}
\baselineskip=12pt
\centerline{\tenit Sezione di Milano, Via Celoria 16, I-20133 Milano, Italy}
\vspace{0.9cm}
\abstracts{ We discuss   a two-dimensional  lagrangian model with $N=2$
supersymmetry  described by  a K\"{a}hler potential $K(X,\bar{X})$
and superpotential $gX^k$ which explicitly exhibits renormalization group
flows to infrared fixed points where the central
charge has a value equal that of the $N=2$, $A_{k-1}$ minimal model. We
consider
the dressing of  such models by N=2 supergravity:
 in contradistinction to bosonic or $N=1$ models, no modification  of the
$\b$-function takes place.
}

\vfil

It has been recognized for some time that $N=2$ minimal models can be viewed as
critical points of Landau-Ginzburg theories, and a considerable body of
literature has developed around this idea
\cite{Kastor,Vafa,Howe,Cecotti,Marshakov,Liao,Witten}.
It is generally believed  that  in a field-theoretic language such models
are described, at and away from the fixed points, by $N=2$  superspace actions
of the form
\EQ
{\cal S} = \int d^2x d^4 \th~ K(X, \Xb ) +\int d^2xd^2 \th~ W(X) +\int d^2x
d^2\bar{\th}~ \bar{W}(\Xb)
\EN
where $K(X,\bar{X})$ is the K\"{a}hler potential, function of chiral and
antichiral superfields $X$, $\bar{X}$, while the superpotential $W(X)$
is a quasi-homogeneous polynomial in the chiral superfields. These ideas have
been
tested in numerous ways, but no complete lagrangian models have been
constructed which exhibit explicitly this behavior.

 Away from the fixed points, along the renormalization group  trajectories, the
$N=2$ nonrenormalization theorem ensures
that the form of the superpotential remains unchanged while
the K\"{a}hler potential flows according to quantum corrections  in such a way
that at
the fixed points the resulting action describes superconformally
invariant systems.  In a complete lagrangian description one would like to
exhibit a suitable K\"{a}hler potential such that, for example, for a simple
Landau-Ginzburg superpotential $gX^k$ the system flows to an infrared fixed
point
where the central charge is the one of the $A_{k-1}$ N=2 minimal model. We
present here
such a  K\"{a}hler potential.  Generalizations to other minimal models are
straightforward.\cite{flows}

We first
examine the situation at the fixed point. In the absence of the superpotential
$W$, a
generic $N=2$ $\s$-model is classically (super)conformally invariant. In the
presence of the
superpotential the stress-energy tensor (the supercurrent actually) acquires a
classical
trace (a supertrace) and for a general $K(X,\bar{X})$ no improvement term
 can be found to make the
theory superconformally invariant. In fact, for a given superpotential, the
condition
of scale invariance fixes   the K\"{a}hler potential  and the improvement
term up to a normalization factor.
As we shall see, it is this normalization factor that determines the central
charge, and it
is a specific normalization factor that gets selected,  when we start away from
the critical point, by the renormalization group flow.  (The critical point
form of our lagrangian has been
also described by Marshakov \cite{Marshakov}
and, in its Liouville version, by Liao and Mansfield
\cite{Liao}, but in   these
references the normalization factor could not be determined.)

We work in Minkowski space with light-cone variables
\bea
x^{\pp} =\frac{1}{\sqrt{2}} (x^0+x^1) ~~~~,~~~~\pa_{\pp}= \frac{1}{\sqrt{2}}
(\pa_0+\pa_1) \nonumber\\
x^{=} =\frac{1}{\sqrt{2}} (x^0-x^1) ~~~~,~~~~\pa_{=}= \frac{1}{\sqrt{2}}
(\pa_0-\pa_1)
\ena
and
\EQ
\Box \equiv \pa^{\m}\pa_{\m}  =2\pa_{\pp}\pa_{=} ~~~~,~~~~ \pa_=
\frac{1}{x^{\pp} }= 2\pi i \d^{(2)}(x)
\EN
The superspace spinorial
coordinates are $\th^+$, $\th^-$, $\bar{\th}^+$, $\bar{\th}^-$,
and the corresponding covariant derivatives satisfy
\EQ
\{D_+, \bar{D}_+\} = i \pa_{\pp} ~~~~,~~~~\{D_-,\bar{D}_-\} = i \pa_=
\EN
with all other anticommutators vanishing. For a kinetic term
$\int d^2x d^4\theta~ X \Xb $ the
chiral
field propagator is
\EQ
<X(x, \th ) \Xb (0)> = -\frac{1}{2\pi} \bar{D}^2 D^2 \d^{(4)}(\th )
\ln[m^2(2x^{\pp}x^= +\ell ^2)]
\EN
where $m$ and $\ell$ are infrared and ultraviolet cutoffs respectively. We have
defined
$D^2 \equiv D_+D_-$ and $\bar{D}^2 \equiv \bar{D}_+\bar{D}_-$.

We couple the system described by Eq. (1) to
  $N=2$ supergravity and in order to conveniently describe the above-mentioned
improvement of the supercurrent  we include a
chiral  ``dilaton''
term, so that the action takes the form
\bea
{\cal S}&=& \int d^2x d^4 \th E^{-1} K( e^{iH.\pa} X, e^{-iH.\pa }\Xb ) +\int
d^2x d^2 \th e^{-2 \s}
W(X) \\
&+&\int d^2x d^2 \bar{\th} e^{-2\bar{\s}} \bar{W}(\Xb )
+\int d^2x d^2\th~ e^{-2\s}R~ \Psi(X) +
\int d^2x d^2 \bar{\th}~ e^{-2\bar{\s}}\bar{R}~\bar{\Psi}(\Xb)
\nonumber
\ena
Here the vector superfield $H$ and the chiral compensator $\s$ are the
supergravity prepotentials.
At the linearized level we have the explicit expressions \cite{flows}
\bea
E^{-1}&=&1-[\bar{D}_+,D_+]H_=-[\bar{D}_-,D_-]H_{\pp} \nonumber\\
R&=& 4\bar{D}_+\bar{D}_- [\bar{\s} +D_+\bar{D}_+H_=+D_-\bar{D}_-H_{\pp}]
\nonumber\\
\bar{R}&=&4D_+D_-[ \s  -\bar{D}_+D_+H_=  -\bar{D}_-D_-H_{\pp}]
\ena
 The general solution of the constraints of  $N=2$ supergravity is given in
Ref. 9.

The invariance of the supergravity-coupled system under local
supersymmetry transformations is expressed by the conservation law
\EQ
\Db_- J_{\pp}  = D_+J ~~~~,~~~ D_-J_{\pp}= \Db _+\bar{J}
\EN
where
the supercurrent is given by
\EQ
J_{\pp} \equiv \frac{\d {\cal S}}{\d H_=}|_{H,\s =0} =2 [D_+X \Db_+\Xb K_{X\Xb}
-2
\Db_+D_+\Psi +2D_+\Db_+\Psib]
\EN
and the supertrace by
\EQ
J \equiv \frac{\d {\cal S}}{\d \s}|_{H,\s =0}= -2[W -2 \Db_+\Db_-\Psib]
\EN
We have introduced the K\"ahler metric
\EQ
K_{X\Xb}= \frac{\pa^2K}{\pa X \pa \Xb}
\EN

Superconformal invariance requires the supertrace $J$ to vanish. For the
superpotential $W=gX^k$ the equations of motion (with the notation $K_X \equiv
\pa_XK$, etc.)
\EQ
\Db_+\Db_- K_X +W_X=0
\EN
give
 \EQ
W = - \frac{1}{k} \Db_+ \Db_- (X K_X)
\EN
 Using  this expression in Eq. (10), the condition  for superconformal
invariance, $J=0$, $\bar{J}=0$,  requires
\EQ
X K_X = -2k\Psib (\Xb ) ~~~~,~~~~ \Xb K_{\Xb} =-2k \Psi (X)
\EN
modulo a {\em linear} superfield  which gives
no contributions to
the action. We have assumed that  $\Psi$, $\Psib$ are local, and
(anti)chirality and
dimensionality require them to be  functions of  $X$, $\Xb$ respectively.
The equations above can be immediately integrated and give
\bea
K &=& \a \ln X \ln \Xb \nonumber\\
\Psi &=& - \frac{\a}{2k} \ln X ~~~~,~~~~\Psib = - \frac{\a}{2k} \ln \Xb
\ena
with arbitrary constant $\a$. Using the field redefinition $X \equiv
 e^{\Phi}$ the
corresponding lagrangian
can be recast in Liouville form.

We compute now the conformal anomaly  of our  model, and show that the central
charge of the fixed point theory equals the central charge of  $N=2$,
$A_{k-1}$
minimal models when $\a$ is suitably chosen.
It is given by the coefficient in front of  the  induced
supergravity
effective action  $R \Box ^{-1} \bar{R}$  obtained by integrating out the
fields $X$, $\bar{X}$, and  can be determined by  contributions to the
 $H_=$ self-energy, from which the covariant expression can be
reconstructed.

We compute away from the fixed point, using an effective
configuration-space propagator
\EQ
<X(x, \th ) \Xb (x', \th ')> = -\frac{K^{X\Xb}}{2\pi} \bar{D}^2 D^2
\d^{(4)}(\th  - \th ') \ln\{m^2[2(x-x')^{\pp} (x-x')^= +\ell ^2]\}
\EN
where $K^{X\Xb}$ is the inverse of the K\"ahler metric
(cf.  Ref. 12, Eq. (3.13);
additional  terms, involving derivatives of the K$\ddot{\rm a}$hler metric
in the propagator do not
give relevant contributions).  The couplings to $H_=$ can be read from the
 supercurrent in Eq. (9). The relevant vertex from the K\"ahler potential
is
\EQ
2i  \int d^4 \th ~ H_= D_+X \Db_+\Xb K_{X\Xb}
\EN
This leads to  the one-loop contribution
\EQ
-\frac{1}{\pi^2} H_=\frac{\Db_- D_-\Db_+ D_+}{(x-x')^2_{\pp} }H_=
\EN
{}From the dilaton term coupling
\EQ
4 \int d^4 \th (\Psib  - \Psi )\pa_{\pp}H_=
\EN
we have the tree
level contribution
\EQ
-\frac{16}{\pi} \Psi_X K^{X\Xb}\Psib_{\Xb} H_=\frac{\Db_- D_-\Db_+ D_+}
{(x-x')^2_{\pp} }H_=
\EN
Using the relation between the K$\ddot{\rm a}$hler potential and the
improvement term at the fixed point given in Eq. (15), we obtain the final
result
\EQ
-\frac{1}{\pi^2}H_=\frac{\Db_- D_-\Db_+D_+}{(x-x')^2_{\pp}}
H_=\left(1+\frac{4\pi \a}{k^2}\right)\Rightarrow
\frac{1}{4\pi}  R \frac{1}{\Box} \bar{R} \left(1+\frac{4\pi \a}{k^2}\right)
\EN
{}From this expression we obtain the central charge of the system
\EQ
c= 1+ \frac{4\p \a} {k^2}
\EN
For $\a = - \frac{k}{2\pi}$ it equals
   the correct value for the $N=2$, $A_{k-1}$ minimal model,
\EQ
c=1-\frac{2}{k}
\EN

We exhibit now a system which flows in the IR region to
the superconformal theory defined above.  It has two properties: its
$\b$-function is
one-loop exact, so that we can make all-orders statements about the flows; and,
although
it contains arbitrary parameters, the flow to the fixed point uniquely picks
out values which
give the correct central charge for identification with the $A_{k-1}$ minimal
models.
The theory is described by the superpotential
$gX^k$
and the K\"ahler metric
\EQ
K_{X\Xb} = \frac{1}{1+bX\Xb +c (X\Xb)^2}
\EN
corresponding to the K\"ahler potential
\EQ
K = \int dX d\Xb K_{X\Xb}=
 X\Xb  -\frac{b}{4}(X \Xb )^2 +\frac{b^2-c}{9}
(X\Xb )^3+\cdots
\EN

Quantum corrections give rise to divergences that can be reabsorbed
by renormalization of the parameters $b$,
$c$, and
 wave-function renormalization. Actually it is convenient to rescale the
field, $X \rightarrow  a^{-\frac{1}{2}}X$, so that the K\"ahler metric and
superpotential become
(with a redefinition of the parameter $c$)
\EQ
K_{X\Xb} = \frac{1}{a+bX\Xb +c (X\Xb)^2} ~~~~~,~~~~~ g a^{-\frac{k}{2}} X^k
\EN
(A related metric, with $a=c$,
has been discussed  in a bosonic $\s$-model
context by
Fateev {\em et al} \cite{Fateev}. The authors of Ref. 11 have
speculated on the
relevance of such metrics for studying $N=2$ flows.).

Thus as in a standard $\s$-model approach one renormalizes the K\"ahler
metric
 including the parameter $a$  (this is equivalent to  wave-function
renormalization). According to Eq. (26), since the superpotential is not
renormalized, a renormalization of the parameter $a$ leads to a
corresponding renormalization of the coupling
constant $g$.

At the one-loop level the divergence is  proportional to
the Ricci tensor,
\EQ
-(\frac{1}{2\pi} \ln m^2 \ell^2)R_{X\Xb}=(\frac{1}{2\pi} \ln m^2 \ell^2)
\frac{ab+4acX\Xb+bc(X\Xb)^2}{[a+bX\Xb+c(X\Xb)^2]^2}
\EN
so that the K\"{a}hler metric, including one-loop corrections, becomes
\EQ
K_{X \Xb}+ \Delta K_{X \Xb}=
\frac{1}{a(1-\Lambda b)+(b- 4\Lambda ac)X\Xb +c(1-\Lambda a) (X\Xb)^2}
\EN
where $\Lambda \equiv \frac{1}{2\pi} \ln m^2 \ell ^2$.
The original parameters in the
classical lagrangian are then expressed
in terms of renormalized ones:
\bea
a&=&Z_a a_R~~~~,~~~~b=Z_b b_R~~~~,~~~~c=Z_c c_R \nonumber\\
g&=&\m Z_g g_R
\ena
where $\m$ is the renormalization mass scale, and $Z_gZ_a^{-\frac{k}{2}}=1$
as required by the $N=2$ nonrenormalization theorem.
{}From Eq.(28) we find
\bea
Z_a&=& 1+ b(\frac{1}{2\pi} \ln \m^2 \ell^2)\nonumber\\
Z_b&=& 1+\frac{4ac}{b} (\frac{1}{2\pi} \ln \m^2 \ell^2)\nonumber\\
Z_c&=& 1+b(\frac{1}{2\pi} \ln \m^2 \ell^2)\nonumber\\
Z_g&=& 1+\frac{bk}{2}(\frac{1}{2\p} \ln \m^2 \ell^2)
\ena

Defining $t=\ln \m$, the renormalized
parameters satisfy the following renormalization group equations
(in the following we drop the subscript $R$)
\bea
\frac{da}{dt}&=& -\frac{1}{\pi}ab \nonumber\\
\frac{db}{dt}&=& -\frac{4}{\pi}ac \nonumber\\
\frac{dc}{dt}&=& -\frac{1}{\pi}cb \nonumber\\
\frac{dg}{dt}&=&-(1+\frac{b}{2\p}k)g
\ena
Conformal invariance is achieved at the zeroes of the coupling
$\b$-functions. In particular
we are looking for a nontrivial IR fixed point for the
coupling constant $g$, i.e. such that $b(t) \rightarrow -\frac{2\pi}{k}$ as
$t \rightarrow - \infty$. Thus we study the solutions of the system in
Eq. (31) and
select the relevant trajectories.
The equations in (31) have two invariants, the ratio
\EQ
\frac{a}{c}=\r
\EN
and the combination, which we choose to make positive and parametrize suitably,
\EQ
b^2-4ac = b^2-4\r c^2 =(\pi \l)^2
\EN
Here $\r$ and $\l$ are arbitrary constants parametrizing individual
trajectories.

In the $b$-$c$ plane we obtain two types of trajectories, hyperbolas or
ellipses, depending on the
sign of $\r$.
Since we are interested in trajectories
with two fixed points we write  the elliptical solutions,
with $\r <0$.
(The  bosonic model studied in Ref. 10,  written in a different
coordinate
system,  has $\r=1$.)
\bea
b(t) &=& \pi \l  \tanh \l t \nonumber\\
a(t) &=& \pm \frac{\pi \l  \sqrt{-\r}}{2} (\cosh \l t)^{-1} \nonumber \\
c(t) &=& \mp  \frac{\pi \l}{2 \sqrt{-\r}} (\cosh \l t)^{-1}\nonumber\\
g(t) &=& g_0 e^{-t}[\cosh \l t ]^{-\frac{k}{2}}
\ena

The wanted IR fixed point is reached by flowing along trajectories which have
\EQ
\l = \frac{2}{k}
\EN
In this case the superfield $a^{-\frac{1}{2}} X$
acquires anomalous dimension $1/k$ in the
corresponding IR conformal theory, while $a$ and $c$ flow to zero.
Therefore, the effective lagrangian with K\"ahler potential
$K(X,\Xb,a(t), b(t), c(t))$
and superpotential $W(X,g(t),a(t))$ has the following behaviour
 in the infrared,
\EQ
t \rightarrow - \infty~~~~,~~~~
K(t) \rightarrow  -\frac{k}{2\pi} \ln X \ln \Xb  ~~~~,~~~~ W(t)\rightarrow
g_0 X^k
\EN
The improvement term at the IR fixed point has $\Psi = \frac{1}{4\pi} \ln X$.

Changing variables, $X \equiv e^{\Phi}$, leads to the  Liouville lagrangian
\EQ
{\cal L} = -\frac{k}{2\pi} \bar{\Phi}\Phi +g_0 e^{k \Phi}
\EN
with negative kinetic term and {\em with
normalization determined
by the superpotential} (cf.  Refs. 5,6).

We emphasize that  imposing conformal invariance at the one-loop level,
i.e. $R_{XÊ\Xb}=0$, is sufficient to insure the absence of divergences
at higher-loop orders since the Riemann tensor trivially vanishes as well.
Moreover, while in the bosonic or in the $N=1$ supersymmetric theories
the dilaton
term contributes to the metric $\b$-function, in the $N=2$ case
no metric-dilaton mixing occurs due to the chirality of
$\Psi$.
 Thus at the conformal point we obtain exact, all-order results.

The case of two fields, with superpotential $gX^n +g'X^kY^m$
can be treated in similar fashion \cite{flows}. The flows to the
IR fixed point lead to a central charge
which agrees with the results for the various minimal models described
by a two-field Landau-Ginzburg potential.

At the fixed point the model we have discussed is conformally invariant and
therefore
integrable. We have examined its integrability along the flow, by looking for a
higher-spin conserved current. We observe that for the case $k=1$ the
model
reduces to the supersymmetric complex sine-Gordon system studied by
 Napolitano  and Sciuto \cite{Sciuto}, which is indeed {\em classically}
integrable.  In particular, these
authors have shown that a spin 3/2 conserved current exists for their model.
However, for $k>1$ we have been unable to construct a conserved spin 3/2
current.  Although we suspect
that  no conserved current exists  for our models,    integrability along the
flows remains an open  question.

Finally, we discuss effects due to gravitational dressing
\cite {Schmid,Kogan,Ambjorn,Tanii}. It has been shown that for bosonic theories
one-loop $\b$-functions
in the presence of induced gravity  are related to the corresponding ones
computed in the absence of gravitational effects by the universal formula
\EQ
\b_G = \frac{\k +2}{\k +1}\b
\EN
Here $\k$ is the level of the gravitational $SL(2,R)$ current algebra which
can be expressed in terms of the matter central charge as
\EQ
\k+2= \frac{1}{12}[c-13-\sqrt{(1-c)(25-c)}]
\EN
In the semiclassical limit, $c\rightarrow - \infty$, the dressing becomes
\EQ
\b_G \rightarrow (1+\frac{6}{c}) \b
\EN

The above result was obtained by treating induced gravity in light-cone
gauge and
making use of the corresponding Ward identities \cite{Kogan}, or in conformal
gauge
\cite{Ambjorn}
where induced gravity is described by the Liouville action, making use of the
fact that one must distinguish between the scale defined by the fiducial
metric, and the physical scale defined by the Liouville field. It  can  also
be checked explicitly in perturbation theory by performing a  calculation for
the standard bosonic $\s$-model coupled to induced gravity
 in light-cone gauge.

A similar treatment is possible for $N=1$ and $N=2$ induced
supergravity \cite{dressing}.
The argument is particularly simple  in (super)conformal gauge.  The
idea
\cite{Ambjorn,Tanii} is
that,  in the presence of the Liouville mode, the physical scale
gets modified with respect to the standard
renormalization scale. In two-dimensional gravity the only dimensionful
object, which provides the
physical scale, is the cosmological constant term. In light-cone
gauge the cosmological constant is just a $c$-number, so that the usual scaling
is physical and the modifications of the matter $\b$-functions arise through
new divergent contributions due to the gravitational couplings.
In conformal gauge instead the one-loop matter divergence
does not receive gravitational corrections. However in this case
the cosmological term is renormalized and thus it
determines the new physical scale that enters in the computation of the
dressed matter $\b$-functions.

In (super)conformal gauge the computation
can be performed treating on equal footing the bosonic, $N=0$, and $N=1,~2$
induced (super)gravities. In all cases the presence of the Liouville field
$\phi$
in the cosmological
constant term  determines the physical scale $\L$ through
\EQ
\int d^2z d^{2N}\theta  e^{\a_+ \phi} \Leftrightarrow
\int d^2z d^{2N}\theta \Lambda ^{-2s}
\EN
where $ \a_+$ is the positive solution of
\EQ
- \frac{1}{2} \a (\a +Q) =s
\EN
i.e.
\EQ
\a_+=\frac{1}{2} \left[  -Q +\sqrt{Q^2-8s }\right]
\EN
For the various theories of interest one has, from the dimensionality of
$d^2z d^{2N}\th$,  $s= \frac{1}{2}(2-N)$. Thus,
\bea
N&=&0: ~~~s=1 ~~~~~~~~ÊQ=\sqrt{\frac{25-c}{3}} \nonumber\\
N&=&1: ~~~s=\frac{1}{2}~~~~~~~~ÊQ= \sqrt{\frac{9-c}{2}} \nonumber\\
N&=&2: ~~~s=0 ~~~~~~~~Ê Q=\sqrt{\frac{1-c}{2}}
\ena

The (super)gravity modification of one-loop matter $\b$-functions comes from
the
chain rule relating derivatives with respect to the physical scale $\L$
and the
renormalization scale $\m$ in the absence of gravity
\EQ
\b_G=\frac{\pa ln\m}{\pa ln\L} \b= -\frac{2s}{\a_+ Q} \b
\EN

Using the above expressions, one finds for the ordinary gravity case, $N=0$,
the
result in Eq. (38).
 For $N=1,~2$ using also  the expressions for the level $\k$ of the light-cone
supergravity Ka\v{c}-Moody algebra
\bea
N=1:~~~~\k+\frac{3}{2} &=&\frac{1}{8} \left[c-5- \sqrt{1-c)(9-c)} \right]
\nonumber\\
N=2:~~~~~\k+1 &=& \frac{1}{4}(c-1)
\ena
one finds
\bea
N=1:~~~~~~\b_G&=&  \frac{\k+\frac{3}{2}}{\k+1}\b \nonumber\\
N=2:~~~~~~\b_G&=& \b
\ena
These results can be verified perturbatively in light-cone gauge.
Details are presented in a separate publication \cite{dressing}.

We conclude  that there is no supergravity dressing of $\b$-functions in the
$N=2$ case.
We also note that even in the presence of supergravity the nonrenormalization
theorems
continue to hold so that there is no correction to the superpotential.
Therefore, the
 RG flow results discussed above are not modified by the presence of induced
$N=2$
supergravity.

\section{ Acknowledgments} D. Zanon thanks the Physics Department of
Harvard University for
hospitality during the period when some of this work was done.

\end{document}